\documentclass{IEEEcsmag}
\usepackage[utf8]{inputenc}
\usepackage{amsmath, amssymb}

\usepackage{cite}
\usepackage{graphicx}
\usepackage{url}
\usepackage{xcolor}
\usepackage{siunitx}
\usepackage{gensymb}
\usepackage{textcomp}
\usepackage[numbers,sort&compress]{natbib}
\usepackage{amsmath}
\usepackage{bm}
\usepackage{multirow,tabularx}
\usepackage{threeparttable}
\usepackage{makecell}
\usepackage{array}
\usepackage{graphicx}
\usepackage{subfig}
\newcolumntype{P}[1]{>{\centering\arraybackslash}p{#1}}
\newcolumntype{M}[1]{>{\centering\arraybackslash}m{#1}}
\usepackage[colorlinks,urlcolor=blue,linkcolor=blue,citecolor=blue]{hyperref}
\expandafter\def\expandafter\UrlBreaks\expandafter{\UrlBreaks\do\/\do\*\do\-\do\~\do\'\do\"\do\-}
\usepackage{upmath,color}

\jvol{XX}
\jnum{XX}
\paper{8}
\jmonth{Month}
\jname{Publication Name}
\jtitle{Publication Title}
\pubyear{2025}

\begin{document}

\sptitle{Article Type: Description  (see below for more detail)}


\title{MetaDecorator: Generating Immersive Virtual Tours through Multimodality}

\author{Shuang Xie}
\affil{Shopify Inc., Ottawa, ON, K2P 1L4, Canada}

\author{Yang Liu}
\affil{Shopify Inc., Ottawa, ON, K2P 1L4, Canada}

\author{Jeannie S.A. Lee}
\affil{Singapore Institute of Technology, Singapore, 138683}

\author{Haiwei Dong}
\affil{Huawei Canada, Ottawa, ON, K2K 3J1, Canada}


\begin{abstract}\looseness-1
MetaDecorator, is a framework that empowers users to personalize virtual spaces. 
By leveraging text-driven prompts and image synthesis techniques, MetaDecorator adorns static panoramas captured by 360° imaging devices, transforming them into uniquely styled and visually appealing environments. This significantly enhances the realism and engagement of virtual tours compared to traditional offerings. Beyond the core framework, we also discuss the integration of Large Language Models (LLMs) and haptics in the VR application to provide a more immersive experience.
\end{abstract}

\maketitle

\begin{figure}
\vspace{-0.3cm}
\centering
\includegraphics[width=.48\textwidth, trim={3 15 1 0},clip]{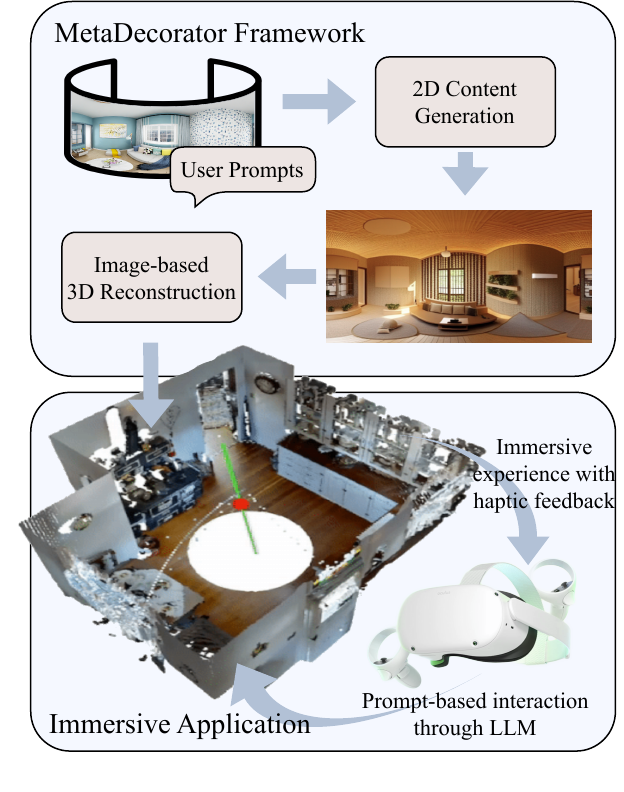}
\caption{The MetaDecorator Framework operates in two primary stages: 1) Image Decoration, where panoramic decorated images are generated based on guiding prompts; and 2) 3D Reconstruction, which produces realistic 3D representations from the decorated images. The framework outputs polygonal meshes, optimizing render speed for edge computing and facilitating integration with VR applications.}
\label{fig:framework}
\end{figure}

\begin{figure*}
\centering
\includegraphics[width=\textwidth, trim={0 3 0 3},clip]{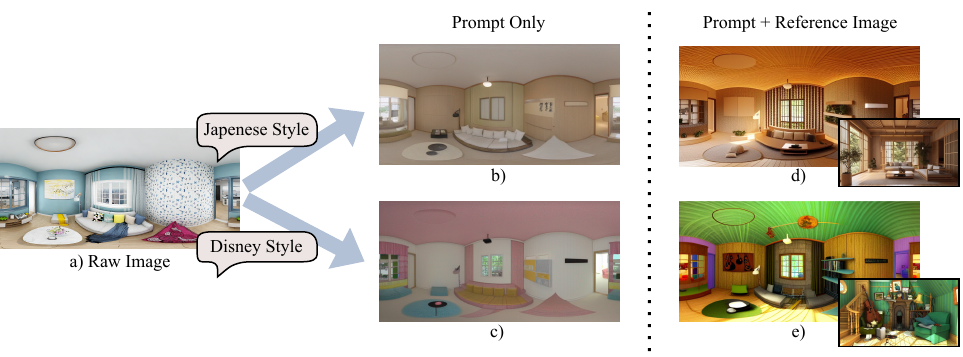}
\caption{Examples of Panoramic Image Decoration: a) Original Image, b-e) Decorated Images. Panels b) and d) showcase Japanese-style decorations, while panels c) and e) feature Disney-style decorations. For panels d) and e), additional reference style images are provided at the bottom right of each respective image.}

\label{fig:example}
\end{figure*}


\chapteri{T}he realm of immersive experiences has undergone a significant transformation with the introduction of 360° virtual tours. These immersive journeys enable users to traverse digitized replicas, or ``digital twins", of actual physical environments, which can be accessed via web browsers and Virtual Reality (VR) headsets \cite{wu2021identifying, 8493267}. Matterport \cite{matterport}, one of the companies leading the charge in this technological revolution, managed 11.7 million spaces encompassing 38 billion square feet by 2023. This achievement signifies the beginning of a new era in digital exploration.


Despite these advancements, current virtual tour technologies exhibit certain constraints. The 3D meshes produced by these systems frequently lack intricate detail and offer limited viewpoints \cite{song2023roomdreamer,li2024panogen}. Moreover, the content captured by camera sensors remains unalterable, providing no scope for customization or personalization. These limitations leave users desiring for more creative control and the ability to modify and stylize their virtual environments unfulfilled.

\section{METADECORATOR FRAMEWORK}


In response to this, our work introduces MetaDecorator, the framework that endows users with the capability to customize their virtual experiences using text prompts. This framework shown in FIGURE \ref{fig:framework} consists of two main stages. To provide the interactive and personalized experience, MetaDecorator first capitalizes on recent advancements in 2D content generation using diffusion models \cite{stablediff} for skybox or panoramic images employed in virtual tours. It utilizes user-inputted text prompts to guide the aesthetic of the synthesized scene or modify the partial region or objects, thereby facilitating the creation of diverse and tailor-made virtual environments.

Unlike conventional methods that treat 3D scenes as disjointed assemblies of multi-view images, MetaDecorator emphasizes the joint optimization of the entire scene to improve the immersive user experience. To do that, MetaDecorator adopts image-based 3D reconstruction with geometry constraints by fully utilizing the high quality RGB images and camera poses available at virtual tours. The Lidar data would be used to improve the reconstructed quality if the virtual tour is obtained by advanced sensors \cite{matterport}. To enable easy integration with VR applications, MetaDecorator outputs polygonal meshes which also facilitate the render speed in edge computation, such as VR devices. With this high extensibility, MetaDecorator can incorporate LLMs to better understand user preferences and interact with the environment in more sophisticated ways, such as automatically furnishing virtual spaces with user prompts. Additionally, integrating haptic feedback into these applications can further enhance the user's sense of presence within the virtual tours.

\section{CASE STUDY: GENERATING IMMERSIVE VIRTUAL TOURS}
Building on the framework outlined earlier, we present a case study demonstrating a pipeline for enhancing indoor panoramic images with user-specified elements and embedding for metaverse applications, allowing for the personalization of virtual tours. The pipeline follows two main stages as shown in FIGURE \ref{fig:framework}: indoor panoramic image decoration and panoramic NeRF rendering. In the first stage, skybox images in the virtual tour applications are converted to panoramic images. Then the panoramic images are partially or fully decorated based on user-provided text prompts using stable diffusion \cite{stablediff}. In the second stage, a NeRF model \cite{mildenhall2021nerf} is used to represent a 360° view of the decorated image into 3D spaces, which is subsequently converted into a 3D mesh object. This mesh object is optimized for efficient rendering, facilitating its seamless integration into metaverse applications.




\subsection{MetaDecorator Keeps Geometric \& Semantic Consistency}
To create a more immersive and realistic decorating experience, MetaDecorator ensures geometric and semantic consistency throughout the 2D content generation process.

Specifically, the 2D content generation begins by taking input from existing virtual tour 360° images and user inputs to generate decorated images. The 360° images are first converted into seamless panoramic representations from skybox images, ensuring style consistency during the decoration process. Depth, edge, and instance segmentation \cite{kirillov2023segment} information are then extracted from the panoramic image, which is crucial for maintaining geometric and semantic consistency in the decorated images. When the user provides a prompt, the system attempts to identify a specific region for decoration based on the semantic segmentation results. If a specific region cannot be identified, the entire image will be decorated. Additionally, users have the option to input a style image as a reference, which enhances control over the artistic outcome by influencing the overall aesthetic direction of the final decorated image.

The 2D content generation process utilizes the inputs mentioned above to create new decorated images that harmoniously blend user-specified decorative elements, artistic style, and the underlying geometric structure of the scene. This is achieved by employing a stable diffusion model in conjunction with ControlNet \cite{zhang2023adding}. ControlNet provides geometric guidance by analyzing depth and edge information to understand the scene's structure. Throughout the iterative refinement process, ControlNet generates intermediate outputs at different resolutions, ensuring that the decorations seamlessly integrate with the existing geometry. The diffusion model incorporates text and style embeddings derived from the user's prompt and reference images, guiding the integration of these elements into the final image while adhering to geometric constraints provided by ControlNet.

The 2D content generation is an iterative process that allows users to gradually adjust the image until it meets their needs. For example, a user can first apply a specific style to the entire scene based on a reference image and then interactively decorate individual elements, such as a sofa, by specifying styles or products based on prompts or additional references. The example of decorated images generated with this process is shown in FIGURE \ref{fig:example}.

\subsection{MetaDecorator Supports Metaverse Applications}

\begin{figure*}
\vspace{-1cm}
\centering
\includegraphics[width=.90\textwidth, trim={0 10 0 0},clip]{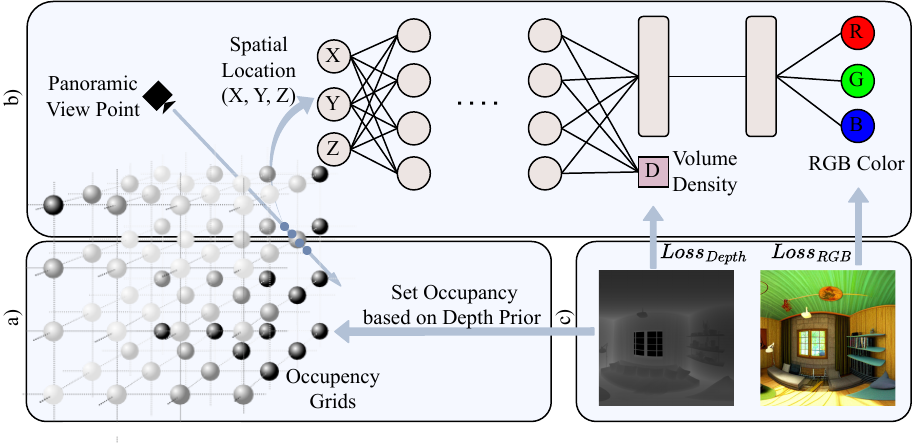}
\caption{The training pipeline of DP-NeRF, optimized for efficiency through the following steps: a) Establishing occupancy based on depth prior; b) Tracing and training the NeRF model exclusively for occupied grids; c) Constraining appearance and geometry by integrating RGB and depth loss.}
\label{fig:nerf}
\end{figure*}

Once the decorated panoramic image is obtained, we propose the Depth Prior and Constraint Panoramic NeRF (DP-NeRF) to reconstruct the 3D space, creating a seamless and immersive experience. DP-NeRF integrates the depth image generated in the previous step with the decorated RGB image to further constrain geometric consistency during the 3D reconstruction process. Additionally, it leverages the depth information to optimize the training process, enhancing both efficiency and quality. The details of this optimization process are discussed in the next section.



Though DP-NeRF marks a significant advancement in image-based 3D reconstruction, their implicit volumetric representations differ from widely used polygonal meshes and lack support from common 3D software packages. To overcome this, a coarse to fine two steps approach to generate the high quality mesh. The coarse geometry is first generated by using the marching cubes algorithm \cite{lorensen1998marching}, which is applied to the ray tracing results derived from the NeRF model. An iterative surface refinement process then adjusts vertex positions and face density based on re-projected rendering errors, refining both appearance and geometry \cite{10377697}. We finally convert the model into the high quality mesh with texture images for real-time rendering. As a result, this method not only preserves rendering quality and performance for green AI, but also facilitates easy integration with common 3D software and metaverse applications.

\section{ENHANCEMENT: TRAINING EFFICIENTLY IMPROVEMENT FOR GREEN AI}

While NeRF is one of the leading approaches for high-quality image-based 3D reconstruction, it is notably time-intensive, often requiring several hours or even an entire day to complete the training for a single decorated panoramic image.
To enhance the efficiency of NeRF training for green AI and improve the user experience, we propose DP-NeRF as illustrated in FIGURE \ref{fig:nerf}. Traditional NeRF training for a single 360° image involves simulating ray tracing for panoramic images using a camera distortion matrix. However, this approach results in many rays tracing into empty space, which significantly slows down the training process. To accelerate the training process, we first establish an occupancy grid based on the method in \cite{muller2022instant}, where the occupancy of each grid cell is initialized using depth priors obtained from previous steps (illustrated in FIGURE \ref{fig:nerf}a). The occupancy of each grid cell is determined by how close the cell is to the nearest 3D point and the viewpoint. If a grid cell is close to a 3D point, it will have a higher occupancy, while a cell farther away will have lower occupancy. Additionally, the influence of the viewpoint on this occupancy is adjusted by a weight factor, which typically falls within a small range.

After constructing the occupancy grid using depth priors, the NeRF model is trained from multiple simulated viewpoints, following the strategy outlined in \cite{gu2022omni}. During both the training and inference phases, only grid cells with high occupancy value are considered for ray tracing, and the occupancy values are dynamically updated throughout the training process. This approach significantly increases the number of effective training points, thereby accelerating both training and inference.

To further enhance training efficiency, we apply constraints on both the RGB and depth outputs (shown in FIGURE \ref{fig:nerf}c)), which introduces stronger depth control and directly informs the NeRF training. The use of depth constraints not only improves the accuracy of the scene reconstruction but also reduces the likelihood of artifacts, leading to a more robust and reliable model.


\subsection{Prototype Performance}
To assess the performance of DP-NeRF, we use the Peak-Signal-to-Noise Ratio (PSNR) metric. PSNR evaluates the quality of the reconstructed image by calculating the mean squared error in logarithmic space, effectively measuring how closely the synthesized image resembles the original reference image. We compute the PSNR between the reference and rendered images at corresponding input image locations to ensure a precise evaluation. We performed a comparative analysis using a representative scene from the Structured3D dataset, comparing our method against OmniNeRF \cite{gu2022omni} and 360FusionNeRF \cite{kulkarni2023360fusionnerf}. 

The results, presented in Table \ref{tab:eval}, demonstrate that DP-NeRF outperforms both OmniNeRF \cite{gu2022omni} and 360FusionNeRF \cite{kulkarni2023360fusionnerf} in terms of PSNR, achieving a 10x reduction in training time. This highlights DP-NeRF's ability to generate realistic views that closely resemble the original features within the dataset, significantly enhancing training efficiency. 

While DP-NeRF excels in rendering high-quality views from decorated panoramic images (Disney \& Japan Deco shown in Table \ref{tab:eval}), it exhibits a slight performance decrease compared to original images. This is primarily due to noise artifacts present in the depth information and pixel values of the generated decorated images. 


DP-NeRF offers a significant advancement in the efficient training of NeRF models for immersive virtual tours. By leveraging depth priors and constraints, it enhances rendering quality and speed, making it a valuable approach for sustainable AI development.

\begin{table}[b]
\caption{Quantitative Evaluation of NeRF Novel View Synthesis}
\begin{tabular}{cccc}
Image       & Model         & epoch & PSNR  \\ \hline
Raw Image   & Nerf \cite{mildenhall2021nerf}          & 200k  & 22.46 \\
Raw Image   & OmniNerf \cite{gu2022omni}      & 200k  & 26.41 \\
Raw Image   & 360FusionNeRF \cite{kulkarni2023360fusionnerf} & 200k  & 28.05 \\
Raw Image   & our (DP-Nerf) & \textbf{20k}   & \textbf{28.36} \\ \hline
Disney Deco & our (DP-Nerf) & 20k   & 27.16 \\
Japan Deco  & our (DP-Nerf) & 20k   & 27.65
\end{tabular}
\label{tab:eval}
\end{table}

\section{FUTURE IMMERSIVE EXPERIENCES}
\subsection{Introducing LLMs for Better User Interaction}
The next frontier in creating immersive user experiences, particularly in AR/VR applications, lies in enabling natural interaction with multi-modal objects. LLMs have revolutionized user experiences in web and mobile applications, making tasks like searching more intuitive and efficient. With thoughtfully designed LLMs, we can transform the quality of generated environments to precisely align with user needs. Moreover, LLMs can unlock new dimensions of interactivity, such as dynamically retrieving 3D models from database or even generating new models based on user prompts. These models can then be seamlessly embedded into the virtual environment, ushering in a new era of customization, realism, and immersive user experiences. Imagine virtual tours where every element can be tailored in real-time, creating a truly engaging and personalized adventure.

\subsection{Adding Haptic Textures for Immersive Experiences}
Tactile sensation is a critical sensation for users when delving into an immersive virtual environment. Specifically, in the reconstructed 3D environments, different objects (such as walls, windows, etc.) and furniture (e.g., sofas, tables, chairs) have various textures when touched, leading to the haptic texture design. The intuition is that when we slide our fingers on the surface, we basically feel the spatial characteristics of the objects by mechanoreceptors (thousands of sensory receptors detecting stimuli such as touch, pressure, and vibration). By simplification, the principle is that if the design the haptic vibration pattern (such as haptic tones in metaphor \cite{interhaptics}) is carefully done, by considering the aforementioned compelling spatial tactile characteristics, we are able to design different haptic textures for the common objects and furniture in virtual decorated environments, including bricks, wood, tiles, etc.


\bibliographystyle{IEEEtran}
\bibliography{MCE.bib}

\begin{thebibliography}{10}
\providecommand{\url}[1]{#1}
\csname url@samestyle\endcsname
\providecommand{\newblock}{\relax}
\providecommand{\bibinfo}[2]{#2}
\providecommand{\BIBentrySTDinterwordspacing}{\spaceskip=0pt\relax}
\providecommand{\BIBentryALTinterwordstretchfactor}{4}
\providecommand{\BIBentryALTinterwordspacing}{\spaceskip=\fontdimen2\font plus
\BIBentryALTinterwordstretchfactor\fontdimen3\font minus \fontdimen4\font\relax}
\providecommand{\BIBforeignlanguage}[2]{{%
\expandafter\ifx\csname l@#1\endcsname\relax
\typeout{** WARNING: IEEEtran.bst: No hyphenation pattern has been}%
\typeout{** loaded for the language `#1'. Using the pattern for}%
\typeout{** the default language instead.}%
\else
\language=\csname l@#1\endcsname
\fi
#2}}
\providecommand{\BIBdecl}{\relax}
\BIBdecl

\bibitem{wu2021identifying}
X.~Wu and I.~K.~W. Lai, ``Identifying the response factors in the formation of a sense of presence and a destination image from a 360-degree virtual tour,'' \emph{Journal of Destination Marketing \& Management}, vol.~21, p. 100640, 2021.

\bibitem{8493267}
Y.~Liu, H.~Dong, L.~Zhang, and A.~E. Saddik, ``Technical evaluation of {H}ololens for multimedia: A first look,'' \emph{IEEE MultiMedia}, vol.~25, no.~4, pp. 8--18, 2018.

\bibitem{matterport}
Matterport, ``{M}atterport {P}ro3 {C}aptuer {S}ystem,'' \url{https://matterport.com/pro3}, [Online; accessed 10-Aug-2024].

\bibitem{song2023roomdreamer}
L.~Song, L.~Cao, H.~Xu, K.~Kang, F.~Tang, J.~Yuan, and Y.~Zhao, ``Roomdreamer: Text-driven 3{D} indoor scene synthesis with coherent geometry and texture,'' \emph{arXiv preprint arXiv:2305.11337}, 2023.

\bibitem{li2024panogen}
J.~Li and M.~Bansal, ``Panogen: Text-conditioned panoramic environment generation for vision-and-language navigation,'' \emph{Advances in Neural Information Processing Systems}, vol.~36, 2024.

\bibitem{stablediff}
R.~Rombach, A.~Blattmann, D.~Lorenz, P.~Esser, and B.~Ommer, ``Proceedings of high-resolution image synthesis with latent diffusion models,'' in \emph{IEEE/CVF Conference on Computer Vision and Pattern Recognition}, 2022, pp. 10\,674--10\,685.

\bibitem{mildenhall2021nerf}
B.~Mildenhall, P.~P. Srinivasan, M.~Tancik, J.~T. Barron, R.~Ramamoorthi, and R.~Ng, ``{NeRF}: Representing scenes as neural radiance fields for view synthesis,'' \emph{Communications of the ACM}, vol.~65, no.~1, pp. 99--106, 2021.

\bibitem{kirillov2023segment}
A.~Kirillov, E.~Mintun, N.~Ravi, H.~Mao, C.~Rolland, L.~Gustafson, T.~Xiao, S.~Whitehead, A.~C. Berg, W.-Y. Lo \emph{et~al.}, ``Segment anything,'' in \emph{Proceedings of the IEEE/CVF International Conference on Computer Vision}, 2023, pp. 4015--4026.

\bibitem{zhang2023adding}
L.~Zhang, A.~Rao, and M.~Agrawala, ``Adding conditional control to text-to-image diffusion models,'' in \emph{Proceedings of the IEEE/CVF International Conference on Computer Vision}, 2023, pp. 3836--3847.

\bibitem{lorensen1998marching}
W.~E. Lorensen and H.~E. Cline, \emph{Marching cubes: A high resolution 3D surface construction algorithm}.\hskip 1em plus 0.5em minus 0.4em\relax Association for Computing Machinery, 1998, pp. 347--353.

\bibitem{10377697}
J.~Tang, H.~Zhou, X.~Chen, T.~Hu, E.~Ding, J.~Wang, and G.~Zeng, ``Delicate textured mesh recovery from {NeRF} via adaptive surface refinement,'' in \emph{IEEE/CVF International Conference on Computer Vision}, 2023, pp. 17\,693--17\,703.

\bibitem{muller2022instant}
T.~M{\"u}ller, A.~Evans, C.~Schied, and A.~Keller, ``Instant neural graphics primitives with a multiresolution hash encoding,'' \emph{ACM Transactions on Graphics}, vol.~41, no.~4, pp. 1--15, 2022.

\bibitem{gu2022omni}
K.~Gu, T.~Maugey, S.~Knorr, and C.~Guillemot, ``Omni-{NeRF}: Neural radiance field from 360 image captures,'' in \emph{IEEE International Conference on Multimedia and Expo}, 2022, pp. 1--6.

\bibitem{kulkarni2023360fusionnerf}
S.~Kulkarni, P.~Yin, and S.~Scherer, ``360fusion{NeRF}: Panoramic neural radiance fields with joint guidance,'' in \emph{IEEE/RSJ International Conference on Intelligent Robots and Systems}, 2023, pp. 7202--7209.

\bibitem{interhaptics}
Interhaptics, ``{H}aptic {C}omposer,'' \url{https://www.interhaptics.com/tech/haptic-composer}, [Online; accessed 26-July-2024].

\end{thebibliography}

\begin{IEEEbiography}{Shuang Xie} {\,} is a Senior Machine Learning Engineer at Shopify Inc., Canada. Her research interests include artificial intelligence, large language models, and computer vision. She is a Member of IEEE, and a Member of IEEE Women in Engineering. Contact her at shuang.xie@ieee.org.
\end{IEEEbiography}

\begin{IEEEbiography}{Yang Liu}{\,} is a Staff Machine Learning Engineer at Shopify Inc., Canada. His research interests include artificial intelligence, multimedia, metaverse, and computer vision. He is a Member of IEEE. Contact him at yang.liu@ieee.org.
\end{IEEEbiography}

\begin{IEEEbiography}{Jeannie S.A. Lee} {\,} is an Associate Professor at the Singapore Institute of Technology. She is the Director of Programmes and Deputy Director, Center for Immersification. Her research interests include immersive technologies, multimedia and human-computer interaction. Contact her at jeannie.lee@singaporetech.edu.sg.
\end{IEEEbiography}

\begin{IEEEbiography}{Haiwei Dong} {\,} is a Principal Researcher at Huawei Canada, and an Adjunct Professor at the University of Ottawa. His research interests include artificial intelligence, multimedia, metaverse, digital twins, and robotics. He is a Senior Member of IEEE, a Senior Member of ACM, and a registered Professional Engineer in Ontario.  Contact him at haiwei.dong@ieee.org.
\end{IEEEbiography}

\end{document}